\documentclass[10pt,journal]{IEEEtran}  

\IEEEoverridecommandlockouts                              

\usepackage{graphicx} 
\newtheorem{theorem}{Theorem}

\newtheorem{assumption}{Assumption}

\newtheorem{proposition}[theorem]{Proposition}
\newtheorem{definition}{Definition}
\newtheorem{conjecture}{Conjecture}
\usepackage{mathtools}
\usepackage{semantic}
\usepackage{algorithm}
\usepackage{algpseudocode}
\usepackage[thinc]{esdiff}
\usepackage{xcolor}
\usepackage{amssymb}
\usepackage{bm}
\usepackage{hyperref}
\usepackage{url}
\usepackage{algpseudocode}

\definecolor{darkgreen}{HTML}{006400}

\newcommand{\x}{\mathbf{x}}

\newcommand{\y}{\mathbf{y}}

\newcommand{\s}{\mathbf{s}}

\title{\LARGE \bf
Strategic and Fair Aggregator Interactions in Energy Markets: Mutli-agent Dynamics and Quasiconcave Games
}

\author{Jiayi Li, Matt Motoki, Baosen Zhang
\thanks{The authors are with the department of Electrical and Computer Engineering at the University of Washington. Emails: \{ljy9712, mmotoki, zhangbao\}@uw.edu}
\thanks{The authors are partially supported by the Washington Clean Energy Institute.}%
}

\begin{document}

\maketitle
\thispagestyle{empty}
\pagestyle{empty}

\begin{abstract}
The introduction of aggregator structures has proven effective in bringing fairness to energy resource allocation by negotiating for more resources and economic surplus on behalf of users. This paper extends the fair energy resource allocation problem to a multi-agent setting, focusing on interactions among multiple aggregators in an electricity market. 

We prove that the strategic optimization problems faced by the aggregators form a quasiconcave game, ensuring the existence of a Nash equilibrium. This resolves complexities related to market price dependencies on total purchases and balancing fairness and efficiency in energy allocation. In addition, we design simulations to characterize the equilibrium points of the induced game, demonstrating how aggregators stabilize market outcomes, ensure fair resource distribution, and optimize user surplus. Our findings offer a robust framework for understanding strategic interactions among aggregators, contributing to more efficient and equitable energy markets.
\end{abstract}

\section{Introduction}\label{sec:introduction}
\begin{figure}[ht]
    \centering
    \includegraphics[width=\linewidth]{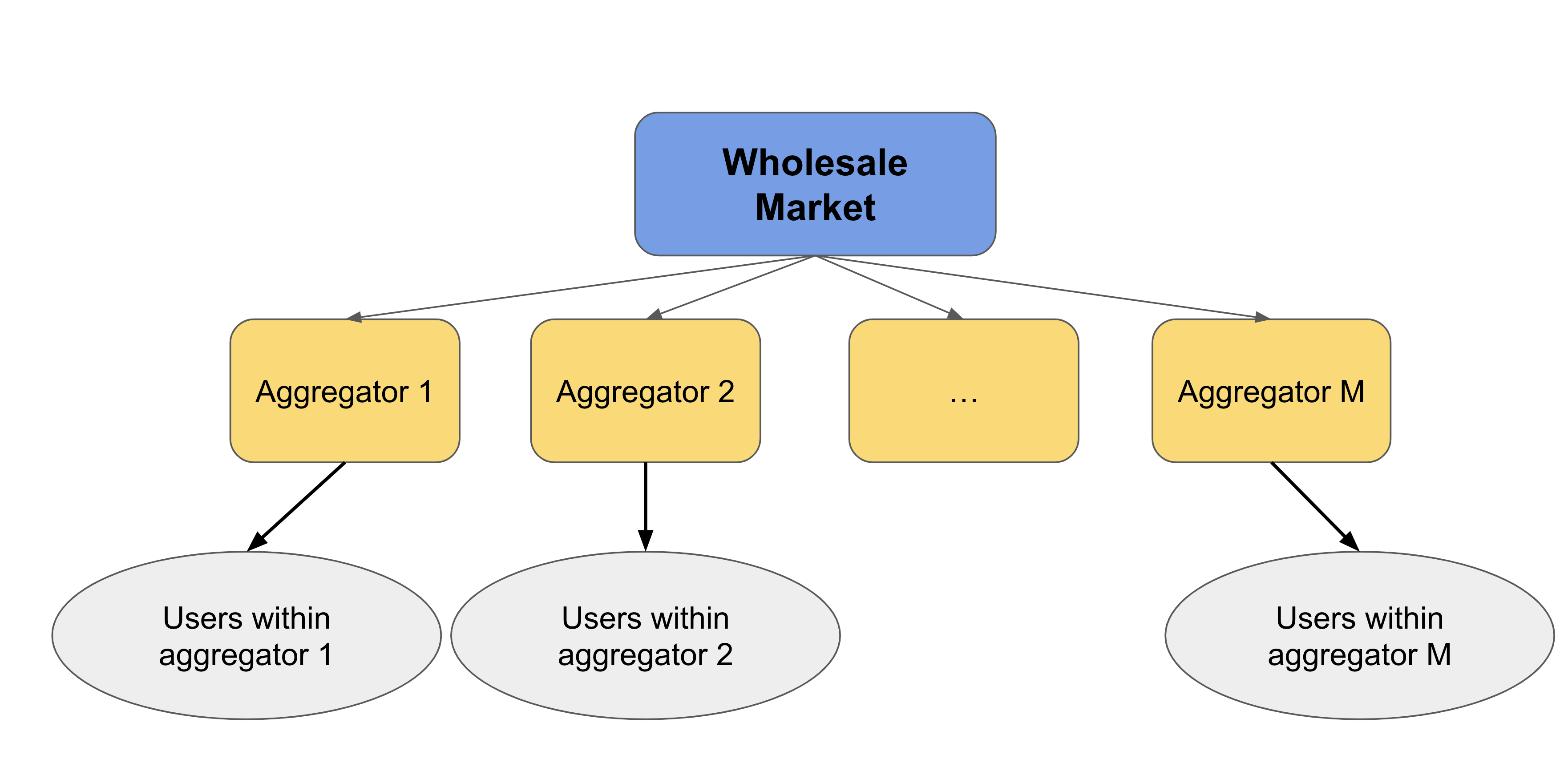}
    \caption{Markets with aggregators can be thought as having a three-layer architecture. The aggregators interact with the market and compete strategically with each other. Within an aggregation, the resources (or benefits) are allocated to each of the users.}
    \label{fig:market_structure}
\end{figure}

In recent years, the barriers for small users to enter the wholesale markets have been successively lowered. On the regulatory side, rulings such as the Federal Energy Regulatory Commission (FERC) Order 2222~\cite{FERC2022} allow users to participate in electricity markets. On the technology side, users have much more flexibility and resources to change their actions to maximize their own benefits. Aggregators play a crucial role in this context by bridging the gap between individual users and the wholesale market~~\cite{burger2017review, chen2023competitive, moret2019energy}. They help mitigate the complexities and transaction costs that individual users would face if they directly participated in the market. At the same time, they reduce the number of agents that a system operator needs to coordinate. Therefore, a system with aggregators can be thought of as having three layers as shown in~Fig.~\ref{fig:market_structure}. In this paper, we address two questions: 1) How should an aggregator allocate resources within its own group and 2) How would different aggregators compete with each other in the market. 

The question of how to allocate resources within an aggregation is becoming more important since the growth of distributed energy resources (DERs) has resulted in more diverse and varied individuals. Existing research often makes the assumption that users within an aggregation are homogeneous or small enough that their differences can be ignored~\cite{zhang2015competition,burger2017review}. However, this assumption may not be valid since users can have drastically different resources and utilities. For example, some may only be able to shift small amount of load, while others may have access to substantial solar and storage. Treating everyone the same could lead to significantly unfair division of benefits, as shown in~\cite{li2024balancing, yang2021optimal,fornier2024fairness}. To mitigate these undesirable consequences, aggregators need to take a more sophisticated approach to how they allocate resources~\cite{li2024balancing}. 

At the level of the aggregators, they simultaneously purchase energy from a wholesale market, where the price of purchasing depends on the total purchase by all of the aggregators. Because the aggregators are (by design) not small entities, we consider the strategic interactions among them. However, moving away from the price taker assumption can lead to challenging technical difficulties since we cannot use the notion of competitive equilibrium, which is the standard solution concept in electricity markets~\cite{kirschen2004fundamental,hobbs2001equilibrium}. In particular, as we shown later in this paper, even under the simplest possible settings, the interaction between the aggregators is not a concave game. In particular, the joint feasible set of actions is not convex, and this rules out a number of game-theoretic tools that are commonly employed to study the electricity market~(see, e.g.~\cite{wei2014competitive,rodriguez2021value,bruninx2019interaction} and the references within). Therefore, it is not easy to characterize whether the interaction between users would even lead to an equilibrium. 


However, the proliferation of distributed energy resources (DERs) has resulted in a more diverse set of user groups ~\cite{akorede2010ders}, challenging traditional allocation schemes that prioritize efficiency. 
 Aggregators' exclusive focus on efficiency can lead to significant disparities in resource allocation and user surplus, as demonstrated in ~\cite{li2024balancing, yang2021optimal,fornier2024fairness}. 
Disparities in the current market structure encourage users to join an aggregator, allowing them to benefit from better-negotiated terms 
~\cite{li2024balancing, sarker2016optimal,contreras2017participation,xie2022information,chen2023competitive}.

\subsection{Summary of Results and Contributions}
This paper extends our previous work \cite{li2024balancing} to a multi-aggregator setting. In~\cite{li2024balancing}, we considered the allocation of energy resources to users within a single aggregator under different fairness metrics. In particular, we showed how the notion of $\alpha$-fairness~\cite{mo2000fair,low2002internet} can be used. In this paper, we show that a pure strategy Nash equilibrium exists  when there are multiple aggregator all strategically competing against each other. 

Showing the existence of a pure Nash equilibrium is nontrivial from the inherent complexities arising from two main factors: the price dependency on total market purchases in the wholesale market, and the need for aggregators to allocate purchased energy to its users while balancing fairness and efficiency. In fact, the resource allocation problem, even when restricted to a single aggregator, is nonconvex. In this paper, through a careful analysis of the optimization problems, we show that while the game between the aggregators is not concave, it is quasiconcave. This allows us to conclude that the game has a pure Nash equilibrium using well-known tools.

Next, we design simulations to demonstrate the interactions among aggregators in the wholesale market and the effectiveness of having an aggregator structure from both the market and the users' perspectives. These simulations provide valuable information on how aggregators can stabilize the market, ensure fair resource distribution, and optimize user surplus. In particular, we show how to balance the need for efficiency and fairness, since neither extreme leads to desirable equilibria. 

The remainder of the paper is structured as follows. Section~\ref{sec:problem-formulation} reviews the fair energy resource allocation problem, introduces the market structures and formalizes the multi-aggregator interaction problems; Section~\ref{sec:game} analyzes the game and provides the main theoretical result; Section~\ref{sec:simulation-results} provides simulation results and interpretations; Finally,  Section~\ref{sec:conclusion-future-work} concludes the paper and offers future research directions.


\subsection{Limitations}
This paper has the following limitations in scope and results. Firstly, we establish the existence, but not the uniqueness of the Nash equilibrium. We believe that this is mostly a technical difficulty, as we conjecture that the Nash equilibrium is, in fact, unique. Secondly, we do not consider how each aggregator is formed nor how stable they are, instead the aggregations are treated as given "as is". We plan to incorporate the work in coalition formations in electricity markets~\cite{pinto2011new,wolff2023dynamic} in future works. Third, the aggregator passes through its revenue and cost to the individual users in this paper. This is consistent with setups where the aggregator takes a fixed "cut" of the profits, but perhaps not with setups where the aggregator delivers a promised payout and can retain the rest of the profits~\cite{lu2020fundamentals}.

\section{Problem Formulation and Preliminaries}\label{sec:problem-formulation}
We assume that the power system market has three layers as shown in Fig.~\ref{fig:market_structure}: an upper-level wholesale market, a middle-level with aggregators, and the lower-level with individual users. This model is fairly standard, with or without middle-level aggregators~\cite{kirschen2004fundamental,burger2017review,chen2024wholesale}.


Suppose that there are $M$ aggregators. For the $j$'th aggregator, suppose that it has $N_j$ users. We denote the energy consumption of the $i$'th user in the $j$'th aggregation as $x_{ji}$. Then the energy consumption of the $j$'th aggregator is $y_j=\sum_{i} x_{ji}$. The wholesale market then determines a price of electricity, $p$, as a function of the consumption of the aggregators $y_1,\dots,y_M$. In this paper, we use $\y=(y_1,\dots,y_M)$ to denote the vector of aggregator consumption, and sometimes write $p(\y)$ when we want to emphasize the dependence of $p$ on $\y$. We also use the standard game-theoretic notation of $\y_{-j}=(y_1,\dots,y_{j-1},y_{j+1},\dots,y_M)$ to denote the consumption of the rest of the aggregators except $j$. 

Given a price $p(\y)$, an aggregator passes this price to its users. Instead of being price-takers as in some studies (see, e.g.~\cite{mathur2017optimal}), we assume users solve their own optimization problems to maximize their utility. That is, a user has a utility function $U$, which represents the value of consuming electricity~\cite{zhang2015competition,zhao2022strategic}. Then the surplus of user $i$ in the aggregation $j$ is defined as 
\begin{equation}\label{eqn:surpus_def}
s_{ji} = U_i(x_{ji}) - p (\y) x_{ji},
\end{equation}
which is a function of its own actions $x_{ji}$ and the system-level consumption $\y$. Next, we will look at the allocation problem within an aggregator and the game between aggregators. 

\subsection{Allocation Problems within an Aggregation}
We think of an aggregator as performing an allocation: given $y_j$, determine the consumption $\{x_{j1},\dots,x_{jN_j}$ for each of the users. Of course, a constraint is that $y_j=\sum_{i=1}^{N_j} x_{ji}$. In addition, the allocation should be optimal in some sense, given the surplus of the users. Let $\s_j=(s_{j1},\dots,s_{j N_1}$, we model the allocation by an aggregator as solving the following optimization problem:
\begin{subequations}\label{eqn:allocation}
\begin{align}
    J(y_j,\y_{-j})=\max_{\x_j} \; & \Phi (\s_j) \\
     \textrm{s.t. } & \sum_{i} x_{ji} = y_j \\
     & s_{ji} = U_i(x_{ji}) - p (\y) x_{ji}, \; \forall{i}\\
     & s_{j} \geq 0 \\
     & \mathbf{x}_j \geq 0.
\end{align}
\end{subequations}
We will go into much more detail about the function $\Phi$ in the next paragraph. We use $J$ as the value of the optimization problem, and it depends on the action of the other aggregators (through the price $p(\y)$). The other constraints define the surplus and restrict the surplus and consumption to be nonnegative. 

The objective function $\Phi$ greatly controls the allocation. The most popular form is the summation of surpluses, defined as $\Phi(\s_j)=\sum_{i=1}^{N_j} s_{ji}$. This is sometimes called social welfare~\cite{mo2000fair,low2002internet},\footnote{This objective has several names, for example, it is sometimes called the total welfare. In this paper, we follow the literature on networking and economics and call it social welfare.} since it maximizes the total (sum) surplus. However, as we have discussed in our prior work~\cite{li2024balancing}, the social welfare tend tend to be unfair, in the sense that it is disproportionately biased towards users with higher utilities. In this paper, we will again use the notion of $\alpha$-fairness, defined as 
\begin{equation}\label{eqn:alpha_fairness}
    \Phi_\alpha (\s_j) = 
    \begin{cases}
        \sum_{i}^{N_j} \frac{s_{ji}^{1-\alpha}}{1-\alpha} &  \text{for $\alpha \geq 0,\, \alpha \neq 1$}, \\
        \sum_{i}^{N_j} {\log{(s_{ji})}} &\text{for $\alpha=1$},
    \end{cases}  
\end{equation}
where we think of $\alpha$ as a nonegative parameter varying from $0$ to $\infty$. When $\alpha=0$, $\Phi_0 (\s_j)=\sum_{i=1}^{N_j} s_{ji}$, recovering the social welfare objective; when $\alpha=1$, $\Phi_1(\s_j)=\sum_{i=1}^{N_j} \log(s_{ji})$, which is the proportional fair objective; and when $\alpha=\infty$, $\Phi_\infty(\s_j)= \min (s_{j1},\dots,s_{jN_j}$, which maximizes the minimum of the surpluses. We interpret $\alpha$ as smoothly trading off efficiency with fairness, with social welfare being the most efficient but the least fair, and MaxMin as the most fair but the least efficient. The notion of $\alpha$-fairness has not been explored in electricity markets, but it has a long history and was first developed in the networking, and interested users can refer to~\cite{mo2000fair,low2002internet}. 

\begin{figure}[ht]   
    \centering\includegraphics[width=\columnwidth]{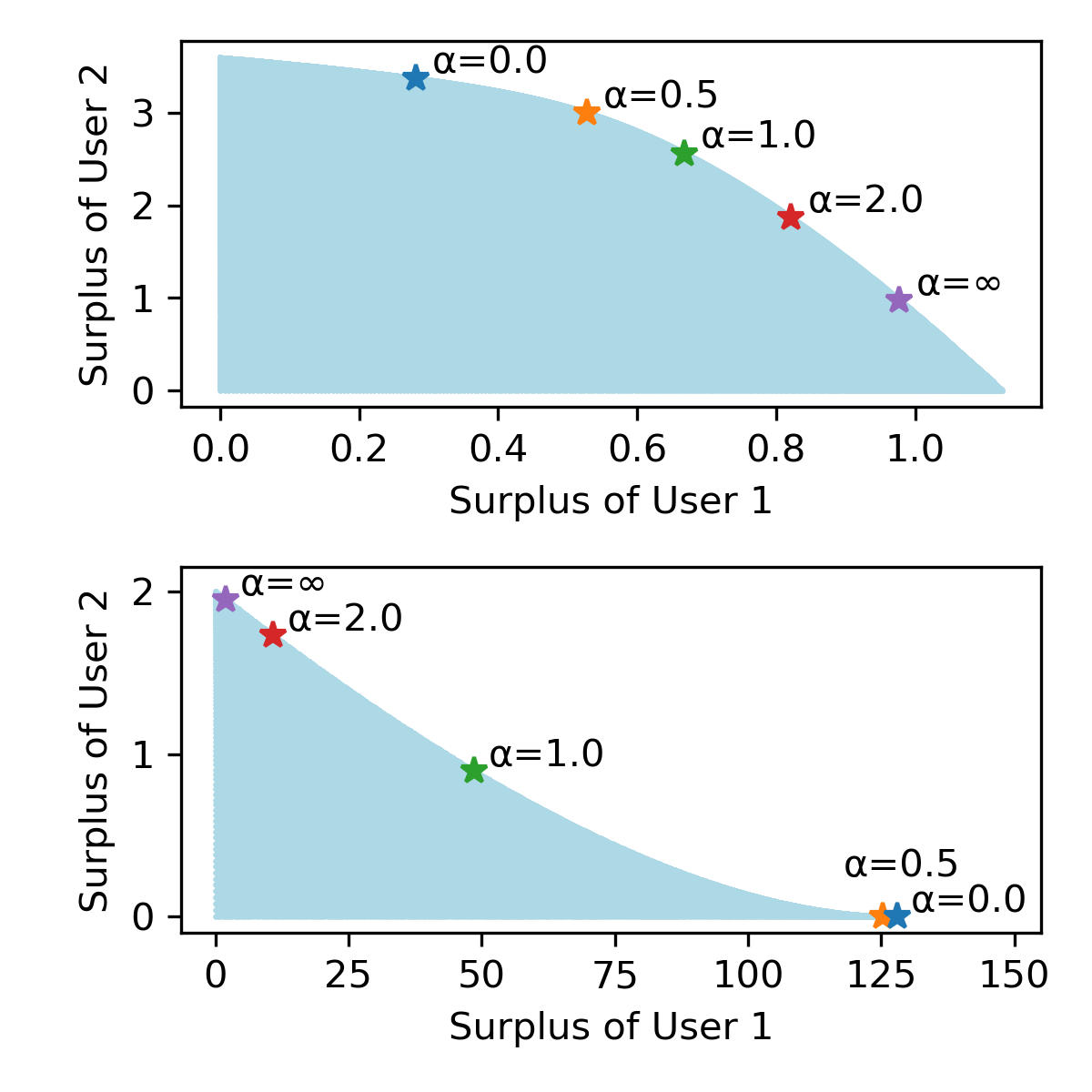}
    \caption{The figure illustrates the feasible regions and Pareto fronts for two-user systems with quadratic utilities. The top panel ($U_1(x_1) = -x_1^2 + 3x_1$ and $U_2(x_2) = -x_2^2 + 6x_2$) shows a convex feasible region, while the bottom panel ($U_1(x_1) = -x_1^2 + 40x_1$ and $U_2(x_2) = -x_2^2 + 4x_2$) shows a non-convex feasible region. Optimal $\alpha$-fairness solutions lie on the Pareto front (the upper right boundary of the feasible region) and increasing $\alpha$ traces out a portion of the Pareto front starting with the least fair social welfare solution ($\alpha=0$) to the most fair max-min solution ($\alpha=\infty$).}
    \label{fig:pareto-front}
\end{figure}

Here we repeat an example in~\cite{li2024balancing} to illustrate some properties of the optimization problems in~\eqref{eqn:allocation}. Consider two users in an aggregation where the total consumption is fixed. The aggregator solves \eqref{eqn:allocation} with respect to different values of $\alpha$. Figure~\ref{fig:pareto-front} shows two examples of the underlying feasible surplus values and how different $\alpha$ explores the Pareto-front. In particular, even for simple utility functions, the underlying feasible sets can be nonconvex. 

\subsection{Game Between Aggregators}
In this paper, we assume that each aggregator is running a fair allocation optimization problem with some $\alpha$, although not necessarily all the same. They naturally form a game, where the $j$'th aggregator's action is $y_j$, and its payoff is $J_j$. Naturally, we restrict $y_j$ to be nonegative. This payoff depends on the actions of all other players because of the price depends on the actions of all aggregators. We are interested in the (pure) Nash equilibria of the game, given by a point $\y^{*}$ where
\begin{equation}\label{eqn:nash}
J_j (y_j^{*},\y_{-j}^{*}) \geq J_j (y_j,\y_{-j}^{*}), \; \forall \; y_j \geq 0, \; \forall \; j=1,\dots,M. 
\end{equation}
At these points, no aggregator can increase its payout by unilaterally changing its action. 

Because the payoffs are value of optimization problems in \eqref{eqn:allocation}, it is not clear what its equilibrium behavior is. For example, we cannot directly use some of the standard game theoretical tools to conclude whether the game has a unique equilibrium, isolated (pure) equilibria, or whether any equalibria exist. This is the case even for simple prices. 

For example, suppose that $p(\y)$ is the sum of $y_j'$: $p(\y)=\sum_j y_j$. Then the payoff of the $j$'th aggregator, with the objective being social welfare (i.e., $\alpha=0$), is 
\begin{subequations} \label{eqn:J_example}
\begin{align}
    J_j(y_j,\y_{-j})=\max_{\x_j} \;  & \sum_{i=1}^{N_j} U_{ji}(x_{ji})-\left(\sum_j y_j \right) x_{ji} \\
     \textrm{s.t. } & \sum_{i} x_{ji} = y_j \\
     & U_i(x_{ji}) - p (\y) x_{ji} \geq 0 \\
     & \mathbf{x}_j \geq 0.
\end{align}
\end{subequations}
Because $\y$ appears in the objective as a product and is also in the constraints, $J_j$ appears to have a complicated dependence on actions $\y$. In the next section, however, we show that we can get a good understanding of the equilibrium of the game through a careful analysis of the allocation problems. 

\subsection{Limiting case: each user being its own aggregator}
When the number of aggregator increases to the point that it is equal to the size of the users, each user would be its own aggregator and therefore fairness-efficiency trade-off turn into its surplus optimization. The formulation for the aggregator $j$ is detailed as: 
\begin{subequations} 
\begin{align}
   \max_{\y_j} J_j(y_j,\y_{-j})= \;  & U_{j}(y_{j})-p( \y)  y_{j} \\
     \textrm{s.t. } & U_{j}(y_{j})-p( \y)  y_{j} \geq 0\\
     & \y_j \geq 0
\end{align}
\end{subequations}
$p(\y)$ is a function on the sum of $\y$.

\section{Nash Equilibria of the Aggregator Game}\label{sec:game}
In this section, we study the Nash equilibria of the game between the aggregators.  We do this by using the following classical result on quasiconcave games:
\begin{proposition}[Existence of Nash Equilibrium for Quasiconcave Games]
    Given $M$ players, with actions $y_1,\dots,y_M$ taking values in compact intervals. Let $J_1(\y),\dots,J_M(\y)$ be the payoff functions. Then if for any player $j$, $J_j$ is quasiconcave in $y_j$ for any fixed values of $\y_{-j}$, then the game has a pure Nash equilibrium.   
\end{proposition}
This proposition dates back to the work in~\cite{debreu1952social,osborne1994course,berry2013economic} and can be stated in much more general forms than the above proposition. We note that it is often given in terms of concave games, but in this paper we do need the more general notion of quasiconcavity. 

We make several standard assumptions on the utility functions of users and the price in the system. 
\begin{assumption} [Utility Function and System Price] \label{ass:Up}
\begin{enumerate}
    \item [a)] The utility functions $U_{ji}(x_{ji})$ are concave for all users and $U_{ji}(0) = 0$.
    \item [b)] The system price $p(\y)=p(\sum_j^{M} y_j)$ and $p(\cdot)=C'(\cdot)$ for some convex function $C$. Equivalently, $p$ is an continuous and increasing function.  
\end{enumerate}
\end{assumption}
The concavity assumption is standard and $U(0)=0$ is a convenient way to ``center'' the utility functions. The assumption on $p$ being an increasing function of total consumption is also a natural assumption. We do note that in this paper the prices are restricted to depend on the sum of aggregators' consumption. This models the case where the system solves an economic dispatch problem to find the electricity price in the system~\cite{kirschen2004fundamental}. In this paper, we do not consider the impact of constraints, although we believe that they do not present fundamental difficulties, and we will consider them in future work. 

We state the main result in the following theorem:
\begin{theorem} \label{thm:main}
Consider the game between the aggregators with payoff functions $J_1(\y),\dots,J_M(\y)$, defined using \eqref{eqn:allocation}. Suppose Assumption~\ref{ass:Up} holds. Then $J_j(y_j,\y_{-j})$ is quasiconcave in $y_j$ for any fixed $\y_{-j}$, and the game has a pure Nash equilibrium.  
\end{theorem}
In fact, we make a stronger conjecture: 
\begin{conjecture}[Uniqueness of the Nash Equilibrium]
With the same setup as in Theorem~\ref{thm:main}, the Nash equilibrium is unique. 
\end{conjecture}
This conjecture is supported by extensive simulation evidence, but we are not able to obtain a rigorous proof.

Due to the fact that the definition of $\Phi_\alpha$ takes on two different forms, we prove Theorem~\ref{thm:main} in two steps, when $\alpha=1$ and $\alpha \geq 0, \alpha \neq 1$. Both would be done through an analysis of the dual multipliers. 

\subsection{Proof of Theorem~\ref{thm:main} when $\alpha=1$}
There are many equivalent ways to define a quasiconcave function. A convenient one for our purposes is 
\begin{definition}
    Consider a continuous function $f(x): \mathcal{I} \rightarrow \mathbb{R}$, where $\mathcal{I}$ is a bounded interval in $\mathbb{R}$. Then $f$ is quasiconcave if there exist a number $x^{*}$ such that $f$ is nondecreasing on $\{x \in \mathcal{I}: x < x^{*}$ and $f$ is nonincreasing on $\{x \in \mathcal{I}: x \geq x^{*}$.\footnote{Note we allow the function to be monotonically increasing or decreasing, where we can take $x^{*}$ to be $\infty$ or $-\infty$.} 
\end{definition}
For more background on quasiconcave functions, the interested reader can consult references such as~\cite{cambini2008generalized}. 

When $\alpha=1$, the payoff function is given by 
\begin{subequations} \label{eqn:J_a=1}
\begin{align}
    J_j(y_j,\y_{-j})=\max_{\x_j} \;  & \sum_{i=1}^{N_j} \log \left(U_{ji}(x_{ji})-p( \y)  x_{ji} \right) \\
     \textrm{s.t. } & \sum_{i} x_{ji} = y_j, \label{eqn:J_a=1_lambda}
\end{align}
\end{subequations}
where we can drop the positivity constraints because of the $\log$ in the objective.  Note that $J_j$ might not be differentiable. But we first assume it is, since it illustrates the main idea of the proof. 

Suppose $J_j$ is differentiable. For a given $\y_{-j}$, consider $y_j$ and $\hat{y}_j$. To show the quasiconcavity of $J_j$, it suffices to show: 1) if $\hat{y}_j >y_j$ and $J_j'(\hat{y}_j,\y_{-j}) \geq 0$, then $J_j'(y_j,\y_{-j})\geq 0$; or 2) if $\hat{y}_j <y_j$ and $J_j'(\hat{y}_j,\y_{-j}) \leq 0$, then $J_j'(y_j,\y_{-j})\leq 0$. These conditions are tantamount to saying that if the function $J_j$ starts to decrease, it cannot increase again. Then the point $y_j^{*}$ required in the definition of quasiconave functions is taken to be when the derivative becomes 0. 

Taking the dual of \eqref{eqn:J_a=1}, we have
\begin{equation}
    \mathcal{L}=\sum_{i=1}^{N_j} \log \left(U_{ji}(x_{ji})-p( \y)  x_{ji} \right)+\lambda (y_j-\sum_{i} x_{ji}),
\end{equation}
where $\lambda$ is the Lagrange multiplier associated with \eqref{eqn:J_a=1_lambda}.
The first order optimality conditions are:
\begin{equation*}
    \frac{U_{ji}'(x_{ji}) -p(\y)}{U(x_{ji})-p(\y)x_{ji}} = \lambda, \; \forall i=1,\dots,N_j,
\end{equation*}
and $J_j'(y_j,\y_{-j})=\lambda$. Now suppose that we have $\hat{y}_j >y_j$, define $\hat{\y}=(\hat{y}_j,\y_{-j}$,  $\hat{\lambda}=J_j'(\hat{y}_j,\y_{-j})$ to be the associated Lagrange multiplier, and $\hat{x}_{ji}$ to the associated primal variables.  

Due to $\sum_i \hat{x}_{ji}=\hat{y}_j>y_j=\sum_i x_{ji}$, and all $x$ and $\hat{x}$ are nonnegative, there exist some $i$ such that $\hat{x}_{ji}>x_{ji}$. Now suppose $\hat{\lambda} \geq 0$, or equivalently,
\begin{equation*}
\frac{U'(\hat{x}_{ji}) -p(\hat{y}_j,\y_{-j})}{U(x_{ji})-p(\hat{y}_j,\y_{-j})} \geq 0.     
\end{equation*}
But by Assumption~\ref{ass:Up}, $U$ is concave and hence $U'$ is nonincreasing, implying that $U'(x_{ji})\geq U'(\hat{x}_{ji})$. Since $p$ is increasing, $p(\hat{y}_j,\y_{-j}) \geq p(y_j,\y_{-j})$. Combining the two, we have $U'(x_{ji}) -p(y_j,\y_{-j}) \geq U'(\hat{x}_{ji}) -p(\hat{y}_j,\y_{-j})$. Since the denominator $U(x_{ji})-p(\y)x_{ji}$ is always positive at optimality, we have $\lambda \geq 0$ if $\hat{\lambda}\geq 0$. This shows that if $\hat{y}_j >y_j$ and $J_j'(\hat{y}_j,\y_{-j}) \geq 0$, then $J_j'(y_j,\y_{-j})\geq 0$. An analogous argument shows that $\hat{y}_j <y_j$ and $J_j'(\hat{y}_j,\y_{-j})\leq 0$, then $J_j'(y_j,\y_{-j})\leq 0$. 

If $J$ is not differentiable, we work with the subgradients, denoted as $\partial_s J_j(y_j,\y_{-j}) = \{s:   J_j(\hat{y_j},\y_{-j}) \geq J_j(y_j,\y_{-j}) + s(\hat{y}_j - y_j)\}$. Since the optimization problem in \eqref{eqn:J_a=1} is convex in $x$,  $\lambda \in \partial_s J_j(y_j,\y_{-j})$. From the above, we have $\hat{\lambda}>0$ implies $\lambda>0$ when $\hat{y}_j > y_j$. Then using the subgradient definition, $J_j(\hat{y_j},\y_{-j}) \geq J_j(y_j,\y_{-j})$, or the function is nondecreasing. Similarly, when $\hat{\lambda}<0$, we have $\lambda<0$, and $\hat{y}_j > y_j$ implies $J_j(\hat{y_j},\y_{-j}) \leq J_j(y_j,\y_{-j})$, or the function is nonincreasing. We can then find the required $y_j^{*}$ required in the definition of quasiconcave functions by selecting a point that has 0 as a subgradient. This finishes the proof. 

\subsection{Proof of Theorem~\ref{thm:main} when $\alpha \neq  1$}

When $\alpha \neq 1$, the payoff function is given by 
\begin{subequations} \label{eqn:J_a>1}
\begin{align}
    J_j(y_j,\y_{-j})=\max_{\x_j} \;  & \sum_{i=1}^{N_j} \frac{\left(U_{ji}(x_{ji})-p( \y)  x_{ji} \right)^{1-\alpha}}{1-\alpha}  \\
     \textrm{s.t. } & \sum_{i} x_{ji} = y_j, \label{eqn:J_a>1_lambda} \\
     &  U_{ji}(x_{ji})-p( \y)  x_{ji} \geq 0, \quad \forall i \label{eqn:J_a>1_sigma} \\
     &  x_{ji} \geq 0, \quad \forall i \label{eqn:J_a>1_mu}
\end{align}
\end{subequations}
where we incorporate positivity constraints and similar to the $\alpha=1$ case, we first assume $J_j$ being differentiable to convey the main idea of the proof. 

Suppose $J_j$ is differentiable. For a given $\y_{-j}$, consider $y_j$ and $\hat{y}_j$. To show the quasiconcavity of $J_j$, it suffices to show: 1) if $\hat{y}_j >y_j$ and $J_j'(\hat{y}_j,\y_{-j}) \geq 0$, then $J_j'(y_j,\y_{-j})\geq 0$; or 2) if $\hat{y}_j <y_j$ and $J_j'(\hat{y}_j,\y_{-j}) \leq 0$, then $J_j'(y_j,\y_{-j})\leq 0$. These conditions are tantamount to saying that if the function $J_j$ starts to decrease, it cannot increase again. Then the point $y_j^{*}$ required in the definition of quasi-concave functions is taken to be when the derivative becomes 0. 

Taking the dual of \eqref{eqn:J_a>1}, we have
\begin{subequations}
\begin{align}
    \mathcal{L} &=\sum_{i=1}^{N_j}\frac{\left(U_{ji}(x_{ji})-p( \y)  x_{ji}\right)^{1-\alpha}}{1-\alpha} \\
    & +\lambda (y_j-\sum_{i} x_{ji}) + \sum_{i=1}^{N_j} [\sigma_i (U_{ji}(x_{ji})-p( \y)  x_{ji}) + \mu_i x_{ji}],
\end{align}
\end{subequations}
where $\lambda$ is the Lagrange multiplier associated with \eqref{eqn:J_a>1_lambda}, $\sigma_i$ is the Lagrange multiplier associated with each \eqref{eqn:J_a>1_sigma} and  $\mu_i$ is the Lagrange multiplier associated with each \eqref{eqn:J_a>1_mu} .
The first order optimality conditions are: 
\begin{equation*}
    \frac{U_{ji}'(x_{ji}) -p(\y)}{\left(U_{ji}(x_{ji})-p(\y)x_{ji}\right)^{\alpha}} - \lambda + \mu_i + \sigma_i \left(U_{ji}'(x_{ji})-p( \y) \right) = 0, \; 
\end{equation*}
\[\forall i=1,\dots,N_j.\] 

For $\x_{ji} > 0$, we have $\mu_i = 0$ by Complementary Slackness and Dual Feasibility of the KKT conditions. In assumption \ref{ass:Up}, we assume that the utility functions $U_{ji}(x_{ji})$ are concave for all users, and $U_{ji}(0) = 0$. Following the assumptions, we have $U_{ji}(0) - 0 * p(\y) = 0$ and $U_{ji}'(x_{ji})$ is decreasing in $x_{ji}$. Moreover, each user's utility is constrained to be positive to achieve non-negative surplus, which indicates that for any feasible $p(\y)$, $U_{ji}' (0)- p(\y) > 0$ and $U_{ji}' (0) > 0$. 

Similarly, without loss of generality, suppose that we have $\hat{y}_j >y_j$, define $\hat{\y}=(\hat{y}_j,\y_{-j})$,  $\hat{\lambda}=J_j'(\hat{y}_j,\y_{-j})$, $\hat{\sigma_i}$ and $\hat{\hat{\mu_i}}$ to be the associated Lagrange multipliers and $\hat{x}_{ji}$ to the associated primal variables.  

Because of $\sum_i \hat{x}_{ji}=\hat{y}_j>y_j=\sum_i x_{ji}$, and all $x$ and $\hat{x}$ are non-negative, there exist some $i$ such that $\hat{x}_{ji}>x_{ji} \geq 0$. By Complementary Slackness, we have $\hat{\mu}_i = 0$.

Compared to the $\alpha = 1$ case, the tricky part is to handle conditions \eqref{eqn:J_a>1_sigma} and \eqref{eqn:J_a>1_mu} and we handle the complexity through case-by-case analysis. 
\subsubsection{Case 1} when $U'(x_{ji}) -p(y_j,\y_{-j}) > 0$, 
$U(x_{ji})-p(\y)x_{ji}$ is increasing as $x_{ji}$ increases.
For $x_{ji} > 0$, $U(x_{ji})-p(\y)x_{ji} > 0$ and the corresponding dual variable $\sigma_i = 0$. 
For $\hat{x}_{ji} > x_{ji} > 0$, we also have $U(\hat{x}_{ji})-p(\y)\hat{x}_{ji} > 0$ with $\hat{\sigma}_i = 0$

\subsubsection{Case 2}
when $ U'(x_{ji}) -p(y_j,\y_{-j}) < 0$,
for $\hat{x}_{ji} > x_{ji} > 0$, 
$0 > U'(x_{ji}) -p(y_j,\y_{-j}) > U'(\hat{x}_{ji}) -p(\hat{y}_j,\y_{-j})$:
$U(x_{ji})-p(\y)x_{ji}$ is decreasing as $x_{ji}$ increases.
Since $U(\hat{x}_{ji})-p(\hat{\y})\hat{x}_{ji}$ is constrained to be positive at optimality, $U(x_{ji})-p(\y)x_{ji} > 0$. The corresponding dual variable $\sigma_i = 0$ and $\hat{\sigma}_i = 0$. \\

In both cases, at optimality point, we have $\lambda \geq 0$, 
\begin{equation*}
    \lambda = \frac{U_{ji}'(x_{ji}) -p(\y)}{\left(U_{ji}(x_{ji})-p(\y)x_{ji}\right)^{\alpha}} \; 
\end{equation*}

Now suppose $\hat{\lambda} \geq 0$, or equivalently,
\begin{equation*}
\frac{U'(\hat{x}_{ji}) -p(\hat{y}_j,\y_{-j})}{(U(\hat{x}_{ji})-p(\hat{y}_j,\y_{-j}))^{\alpha}} \geq 0.     
\end{equation*}
By Assumption~\ref{ass:Up}, $U$ is concave and therefore $U'$ is non-increasing, implying that $U'(x_{ji})\geq U'(\hat{x}_{ji})$. Since $p$ is increasing, $p(\hat{y}_j,\y_{-j}) \geq p(y_j,\y_{-j})$. Combining the two, we have $U'(x_{ji}) -p(y_j,\y_{-j}) \geq U'(\hat{x}_{ji}) -p(\hat{y}_j,\y_{-j})$. 
 This shows that if $\hat{y}_j >y_j$ and $J_j'(\hat{y}_j,\y_{-j}) \geq 0$, then $J_j'(y_j,\y_{-j})\geq 0$. An analogous argument shows that $\hat{y}_j <y_j$ and $J_j'(\hat{y}_j,\y_{-j})\leq 0$, then $J_j'(y_j,\y_{-j})\leq 0$. 

If $J$ is not differentiable, we work with the subgradients, denoted as $\partial_s J_j(y_j,\y_{-j}) = \{s:   J_j(\hat{y_j},\y_{-j}) \geq J_j(y_j,\y_{-j}) + s(\hat{y}_j - y_j)\}$. Since the optimization problem in \eqref{eqn:J_a>1} is convex in $x$,  $\lambda \in \partial_s J_j(y_j,\y_{-j})$. From the above, we have $\hat{\lambda}>0$ implies $\lambda>0$ when $\hat{y}_j > y_j$. Then using the subgradient definition, $J_j(\hat{y_j},\y_{-j}) \geq J_j(y_j,\y_{-j})$, or the function is nondecreasing. Similarly, when $\hat{\lambda}<0$, we have $\lambda<0$, and $\hat{y}_j > y_j$ implies $J_j(\hat{y_j},\y_{-j}) \leq J_j(y_j,\y_{-j})$, or the function is nonincreasing. We can then find the required $y_j^{*}$ required in the definition of quasiconcave functions by selecting a point that has 0 as a subgradient. This finishes the proof.

\section{Simulation Results}\label{sec:simulation-results}
In this section, we present simulation results that illustrate the impact of aggregator structures on market dynamics, particularly focusing on the interactions between large and small users, as well as the interactions among multiple aggregators. The simulations aim to demonstrate how aggregators can enhance market outcomes for small users, especially in the presence of larger users with significant market influence.

For each of the following scenarios, we simulate the long-term multi-agent interactions, including those of aggregators, large-scale users, and small-scale users. Over multiple rounds of market interactions, all participants adjust their strategies according to their optimization schemes. We create visualizations to show the evolution of equilibrium and the convergence of strategies among market participants. Finally, we analyze the outcomes in terms of different types of user surplus and the distribution of total consumption at equilibrium. The code for all simulations is available at: \url{https://github.com/lijiayi9712/aggregator-game}. 

\begin{algorithm}
\caption{Best-Response Dynamics}\label{alg:best-response}
\begin{algorithmic}[1]
\State Initialize the strategy profile $\y = (y_1, y_2, \dots, y_M)$ for all players

\While{$\y$ is not a pure Nash Equilibrium}
    \State For each player $i \in \{1, 2, \dots, H\}:$
    \State Calculate player $i$'s best response $y_i'$ to $\y_{-i}$, 
    \If{$y_i'$ provides a better outcome for player $i$}
        \State Update $y_i \gets y_i'$
    \EndIf
\EndWhile
\State Return the equilibrium strategy profile $\y$
\end{algorithmic}
\end{algorithm}

\subsection{Utility Function of Users}
To perform the simulations, we adopt a quadratic utility for the users, which is commonly used in the literature~\cite{li2017distributed,khezeli2017risk,patnam2021demand}. Other differentiable concave utility functions can be used and do not change the qualitative conclusions. 

More concretely, user $i$ has utility function $U_i(x_i) = -a_i x_i^2 + b_i x_i$, where $a_i$ and $b_i$ are positive numbers. Without other constraints, this function is maximized at $\frac{b_i}{2 a_i}$, and we think of the ratio $\frac{b_i}{a_i}$ as representing the size of the user.

To illustrate how aggregator structure impacts smaller users' performance at equilibrium, we simulate two types of users: small-scale users (e.g. residential users) vs. large-scale users (eg. data centers, industrial factories) whose consumption is large enough to appreciably change the market equilibrium.

For small users, we draw  $a$ and $b$'s from uniform distributions: $a_i \sim \text{Uniform}(0.1, 0.9)$ and $b_i \sim \text{Uniform}(0, 10)$. For larger users, we scale the range of the $b$ parameter so it is much larger than $a$ and use a variable $k$ to denote the magnitude of the market power of the large user. Our goal is to study a market that has both large and small users and to understand how large users would impact the smaller ones. As a general rule, we usually set the large user to be as large as the ``sum'' of all small users. In this regime, the aggregation of users makes a significant difference in their allocations, and we observe the increase in market power of the small users when they are in an aggregate. 


\subsection{Dynamics of the Game}
As proven in the previous section, each aggregator's optimization is quasi-concave, ensuring that the aggregator-aggregator interaction forms a well-defined quasi-concave game. 
We simulate a market with $H$ participants, where each participant decides on the amount of energy to purchase from the wholesale market in each round. The market price is dynamically determined by the total demand of all the market participants. To understand the behavior of the system, we first investigate whether the game would converge to a pure Nash equilibrium under best-response dynamics. 

Best-response dynamics simulate the process when both aggregators and individual users (large and small) play the game repeatedly, and the market dynamics evolve over time as the game progresses through multiple iterations. In the setting, the current purchasing profile $\y$ represents the strategy of each participant in the market: while $\y$ is not a pure Nash equilibrium: There exists at least one user, say user $i$, who can make a beneficial new purchasing strategy $y_i'$, given $\y_{-i}$, that improves its objective value $J_i(y_i', \y_{-i})$ in this round. Therefore, a given strategy $\y$ would lead to a new strategy $\y'$, stopping only at a pure Nash equilibrium.

In Algorithm \ref{alg:best-response}, we summarize the best-response dynamics process, which iteratively updates the purchasing strategies of participants until a pure Nash equilibrium is reached. We note that if a game has multiple Nash equilibria, there is no guarantee to which one the best response dynamics will converge to. However, in all of our simulations, the best response dynamic always converges to the same Nash equilibrium, regardless of how it is initiated. This provides some numerical evidence that the Nash equilibrium is unique (as in Conjecture 1). 

Given the action of the rest of the aggregators, aggregator $i$ solves 
\begin{subequations} \label{eqn:yx}
\begin{align}
    \max_{y_i,\x} \; & \Phi (\s) \\
     \textrm{s.t. } & \sum_{j} x_j = y_i \\
     & s_j = U_j(x_j) - p (y_i,\y_{-i}) x_{j}, \; \forall{j}\\
     & s_{j} \geq 0 \\
     & {x}_j \geq 0,
\end{align}
\end{subequations}
where the optimization is jointly over $y_i$ and the allocations $\x$. This optimization problem is not jointly convex in $y_i$ and $\x$. However, it can be easily solved in a two-step process. Firstly, given $y_i$, the problem is convex in $\x$ and can be solved efficiently. Next, because of the quasiconcavity of the objective value with respect to $y_i$, we can use a simple bisection method to optimize over $y_i$~\cite{agrawal2020disciplined}. Essentially, we can discretize the search space and only need to check a logarithmic number of points in the search space. Together, \eqref{eqn:yx} can be efficiently solved.

We consider the following simulation settings:
\subsubsection{Baseline: Small Individual Users}
We first simulate a scenario in which only small users are present. Each user interacts with the market independently and we analyze the resulting equilibrium in terms of energy consumption, market price, and economic surplus. This is the limiting case where each aggregator has only one user. This baseline setup serves as a reference point for understanding how the presence of a large user and the introduction of aggregators impact market dynamics. 

\begin{figure}[ht]
\centering
\includegraphics[width=0.9\columnwidth]{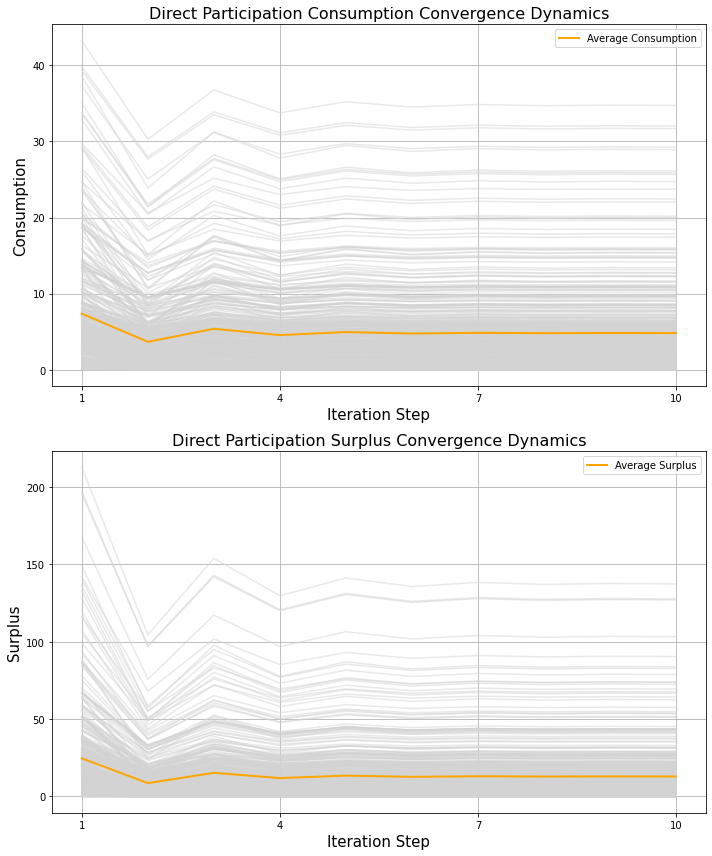}
\caption{Evolution of small users' consumption and surplus under direct market participation: the top plot illustrates the convergence of consumption level over multiple iteration steps, while the bottom plot shows the convergence of surplus. Both plots display the results for 400 small users directly participating in the market. The figures demonstrate how the consumption and surplus dynamics quickly stabilize, converging to a market equilibrium.}
    \label{fig:400-small-market-dynamics}
\end{figure}

We adopt the linear pricing model, that is, $p(\y)=c \left(\sum_j y_j\right)$, for some constant $c$ ($c=0.001$ in our simulations). Linear prices are commonly used in the literature~\cite{zhang2015competition} and other price functions can be used, as long as they are increasing, the qualitative conclusions in this section do not change.  We simulate a system of 400 small users. Figure~\ref{fig:400-small-market-dynamics} shows the evolution of the consumption level (the $x_i$'s) and the surplus of the users. They quickly converges to an equilibrium after a few iterations. 

\subsubsection{multi-aggregator interaction}

\begin{figure}[H] \centering\includegraphics[width=0.8\columnwidth]{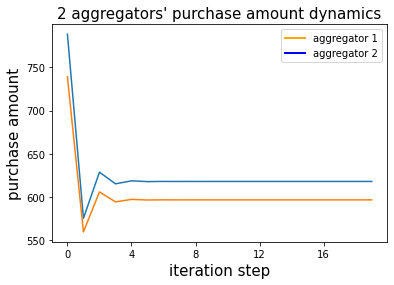}
    \caption{Convergence of strategic purchasing amounts for two aggregators using best-response dynamics: This figure illustrates the convergence of purchase amounts for two aggregators, each representing around 100 small users, as they interact with one large user in the market. The aggregators adjust their strategies iteratively using best-response dynamics, and the market dynamics converge smoothly to an equilibrium point within a few iteration steps. The rapid convergence demonstrates the efficiency of best-response updates in achieving equilibrium in this multi-aggregator setting.}
    \label{fig:market-dynamics-hybrid-agg}
\end{figure}

Figure \ref{fig:market-dynamics-hybrid-agg} presents an example of 2 aggregators playing iteratively: the market dynamics converge smoothly and stably to an equilibrium point very fast through the best-response dynamics update. 




For the rest of the simulation section, we focus on characterizing the converged market equilibrium, explore how aggregator can help the users within, and analyze how aggregators' structure, including their fairness-efficiency trade-offs, competition among aggregators, and the quantity of aggregators impact the resource allocation in the market.

\subsection{Impact of Large Users}
Next, following the baseline system that consists of only small users, we introduce a large user in the system: the large user directly competes with the small users. First, we consider the setting where the small users do not form any aggregation and there is one large user. This large user has a significant consumption level that influences the market dynamics, introduces disparities and affects the equilibrium results for the remaining small users. We compare the market's resulting surplus distribution, in the presence of a large user of different scales, to highlight the disparities that arise when a large user operates without aggregation.

\begin{figure}[H] \centering\includegraphics[width=0.5\textwidth]{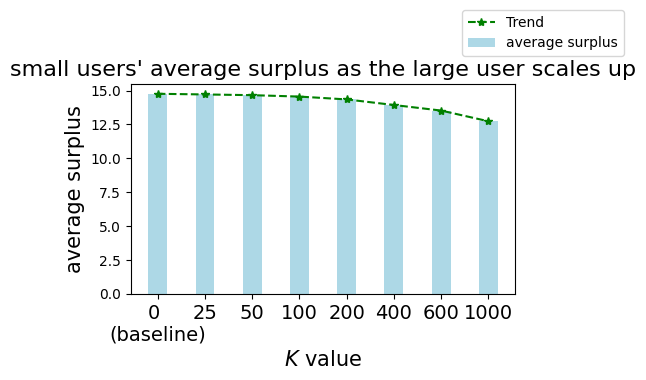}
    \caption{This figure illustrate how the average surplus of the $N$ small users change as the magnitude of the large user increases by changing the value $M$ from 0 to 400. The x-axis is the $K$ value that signifies  and y-axis represents the average surplus amount. We observe that the average surplus decreases as $M$ increases.}
    \label{fig:avg_surplus-increasing-M-no-agg}
\end{figure}

 To investigate how the magnitude of the large user impacts small users at the  market equilibrium, we increase the variable $K$ from 0 to 400 to adjust the size of the single large user that, together with the $N=200$ small users, directly participate in the market, and presents a bar-plot in Fig.~\ref{fig:avg_surplus-increasing-M-no-agg}.
 More concretely, the large user has the utility function $U(x) = -a x^2 + b x$, where $a$ and $b$ are positive numbers. Without other constraints, this function is maximized at $\frac{b}{2 a}$, and we set $K = \frac{b}{a}$ as the measurement of the market power. 
 
Next, we look at what happens when small users participate in the market through joining 2 aggregators that negotiate with the market on their behalf. Each aggregator aims to choose the total amount of consumption that maximizes the $\alpha$-fairness objectives. We then simulate the market interactions among these two aggregators with the large user. Using different fairness measures (e.g., social welfare, proportional fairness, max-min fairness) in the aggregators' decision-making, the aggregators make different purchasing and allocation decisions under each fairness measure, leading to different market outcomes.


\begin{figure}[ht] \centering\includegraphics[width=\columnwidth]{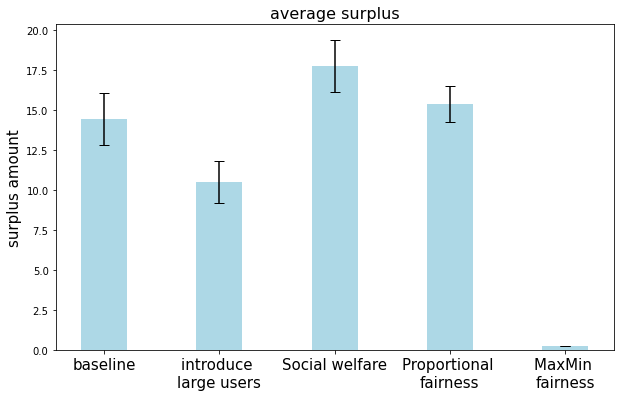}
    \caption{Average surplus for small users under different market configurations and fairness objectives: This figure compares the average surplus achieved by small users across various scenarios: baseline (200 small users), introduction of large users and without aggregators (with the large user having market power equivalent to 200 small users), and three fairness objectives applied within aggregators—social welfare, proportional fairness, and max-min fairness. The bars represent the average values obtained from 50 simulation runs. The results highlight how the introduction of large users and the choice of fairness objectives impact the distribution of surplus among small users, with social welfare objective leading to the highest surplus.}
    \label{fig:surplus-diff-fairness}
\end{figure}

\begin{figure}[ht] \centering\includegraphics[width=\columnwidth]{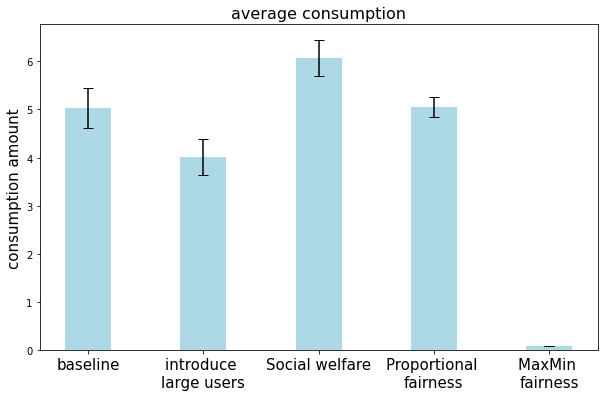}
    \caption{
    Average consumption for small users under different market configurations and fairness objectives under the same setup as in~Fig.~\ref{fig:surplus-diff-fairness}. The results demonstrate the impact of market configurations and fairness objectives on energy consumption, with social welfare and proportional fairness leading to higher consumption levels, while max-min fairness results in significantly lower consumption.}
    \label{fig:consumption-diff-fairness}
\end{figure}

As demonstrated in the comparison bar-plots \ref{fig:surplus-diff-fairness} and \ref{fig:consumption-diff-fairness}, adopting social welfare leads to the highest average surplus for smaller users, with proportional fairness leading to slightly lower values, and MaxMin fairness giving very small surplus values. A natural question is whether an aggregator would ever choose an allocation scheme other than maximizing social welfare, or put in another way, would a user ever join an aggregation that does not use social welfare? 

\begin{figure}[ht] \centering\includegraphics[width=\columnwidth]{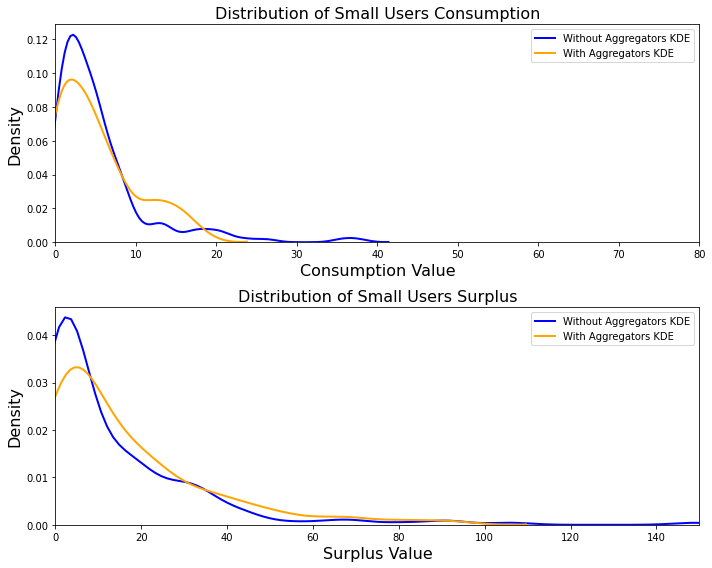}
        \caption{Comparison of Small Users' Consumption and Surplus Distributions With and Without Aggregators optimizing social welfare. The top plot shows the smoothed kernel density estimates (KDE) of small users' consumption values, while the bottom plot displays the KDE of their surplus values. Both distributions are compared under two scenarios: without aggregators (blue) and with aggregators which are optimizing social welfare (orange). The distribution under social welfare is more concentrated, although it still has a long tail.}
    \label{fig:histogram-small-distribution-comparison-SW}
\end{figure}

To answer this question, we look at the distribution of the surplus and consumption of small users. Fig~\ref{fig:histogram-small-distribution-comparison-SW} shows how they change by comparing no aggregation and an aggregator that uses social welfare for allocation. With an aggregator, the distribution of the consumption is more spread out, although most of the users still receive relative little allocation, and the distribution has a long tail, showing that a few users are much better off than others. 

\begin{figure}[ht] \centering\includegraphics[width=\columnwidth]{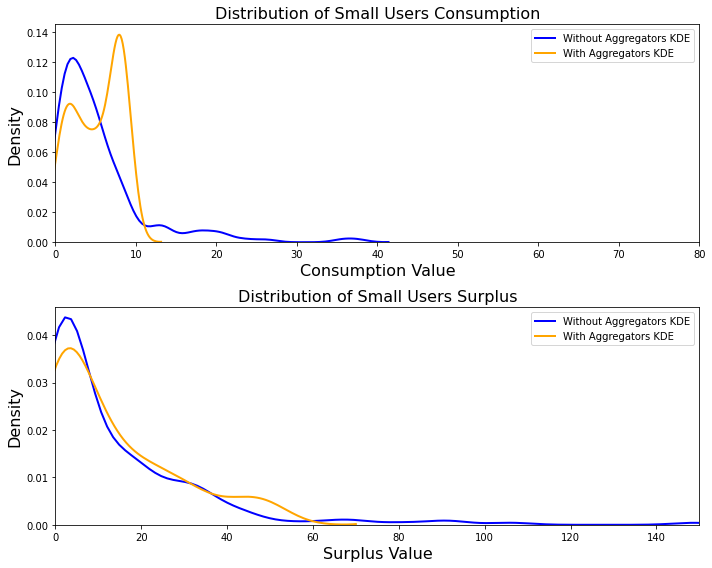}
    \caption{Comparison of Small Users' Consumption and Surplus Distributions With and Without Aggregators optimizing proportional fairness: The top plot shows the smoothed kernel density estimates (KDE) of small users' consumption values, while the bottom plot displays the KDE of their surplus values. Both distributions are compared under two scenarios: without aggregators (blue) and with aggregators which are optimizing proportional fairness (orange). The distribution under proportional fairness is much more uniform, with most values concentrating around the mean.}
    \label{fig:agg-no-agg-comparison-small}
\end{figure}

Fig.~\ref{fig:agg-no-agg-comparison-small} shows the distribution of users' consumption and surpluses when the aggregators use a proportional fair allocation. Here, the distribution is much more ``uniform'', with most users' consumption concentrated around the averaged value and a much shorter tail. Therefore, both social welfare and proportional fair-based allocation have their merits, and users might join an aggregator with these schemes depending on their preferences.

\subsection{Fairness and Competition between Aggregators}
We also investigate how different fairness measures affect the competition between aggregators. That is, what happens when different aggregators run different fairness schemes. We consider two aggregators in the market, and tune the $\alpha$ values of one or both of them from $0$ to $\infty$.

\begin{figure}[ht] \centering\includegraphics[width=\columnwidth]{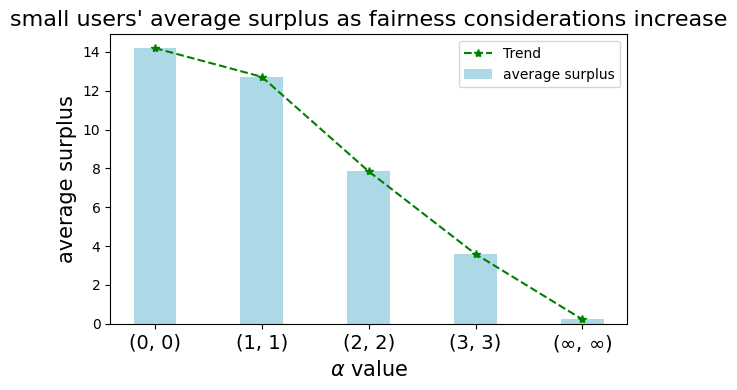}
    \caption{When both aggregators increase their fairness consideration in resource allocation schemes, small users' average surplus decreases.}
    \label{fig:small-surplus-both-increasing_alpha}
\end{figure}

\begin{figure}[H] \centering\includegraphics[width=\columnwidth]{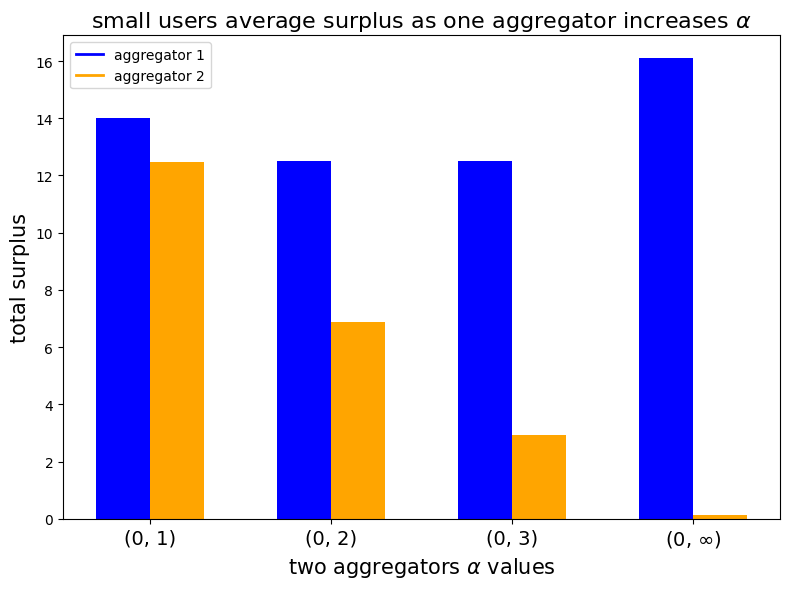}
    \caption{When the first aggregator's resource allocation scheme stays the same, while the second aggregator increases its fairness considerations, the more fairness second aggregator considers, the smaller the average surplus that users within the second aggregators would experience.}
    \label{fig:small-surplus-one-increasing_alpha}
\end{figure}

First, Fig.~\ref{fig:small-surplus-both-increasing_alpha} shows how the average surplus of users changes when both aggregators have the same $\alpha$ and the value of $\alpha$ increases. As $\alpha$ increases (the allocation with an aggregate becomes more fair), the average surplus decreases. Fig.~\ref{fig:small-surplus-one-increasing_alpha} shows what happens when one aggregator holds its $\alpha$ at 0 while the other increases its $\alpha$ value. It shows that the $\alpha$ values should not be too different between the aggregators. These simulations raise an interesting question about how the $\alpha$ values should be set in practice, which we do not answer in this paper but is an important issue to be addressed in the future. 


The results show the user surplus under each fairness measure, the total system surplus and its distribution, and the fairness metrics. 
The results will be visualized in plots showing the user surplus distribution under different fairness measures, the total system surplus versus the fairness measure parameter, and the fairness index versus fairness measure parameter.

\subsection{Impact of Number of Aggregators}
\begin{figure}[ht] \centering\includegraphics[width=\columnwidth]{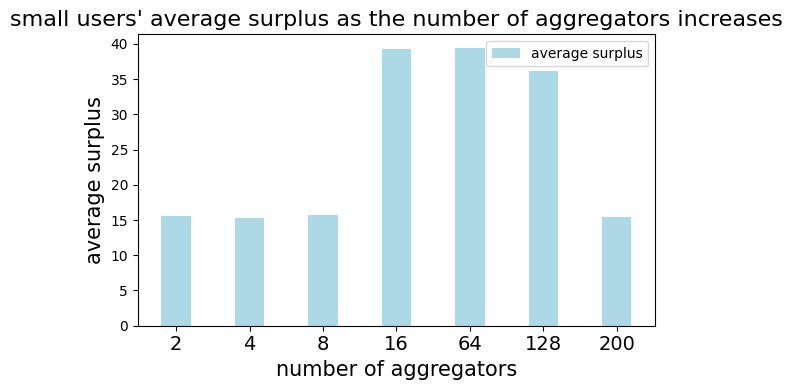}
    \caption{How the number of aggregators impact the small users average surplus at equilibrium when the aggregators are optimizing {social welfare}.}
    \label{fig:increasing_aggregator_surplus_barplot_SW}
\end{figure}

\begin{figure}[ht] \centering\includegraphics[width=\columnwidth]{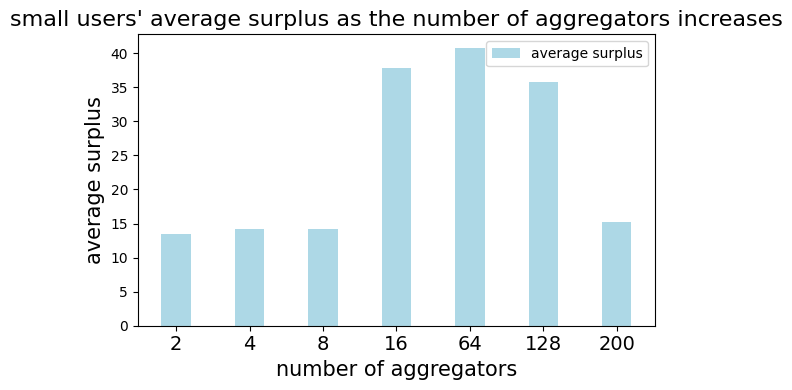}
    \caption{How the number of aggregators impact the small users average surplus at equilibrium when the aggregators are optimizing {proportional fairness}.}
    \label{fig:increasing_aggregator_surplus_barplot_PF}
\end{figure}

Understanding the effect of the number of aggregators on market outcomes is essential for optimal market design and regulatory oversight. To explore this, we conducted simulations where the number of aggregators in the market varied (e.g., 2, 4, 8, ..., N), while keeping their fairness objectives constant (set to be social welfare or proportional fairness for all aggregators). These simulations offer valuable information on how the number of aggregators influences the distribution of average surplus and consumption among small users at the equilibrium point.
Where large users are present, the number of aggregators has a noticeable impact on small users' average surplus, as demonstrated in Fig.~\ref{fig:increasing_aggregator_surplus_barplot_SW} for the social welfare objective and Fig. \ref{fig:increasing_aggregator_surplus_barplot_PF} for the proportional fairness objective. Although no clear monotonic trend is observed as the number of aggregators increases, the data suggest a trade-off between small users' surplus and consumption at the market equilibrium point. This trade-off is particularly relevant for market operators and regulators, who must balance efficiency with fairness while considering the average consumption upper limit imposed by generation capacity. 

For market operators, these results imply that the optimal configuration of the aggregators is context dependent and may require local market experimentation to identify. The number of aggregators can influence market dynamics in ways that are not immediately apparent, affecting both the stability of market prices and the distribution of economic benefits among users. Therefore, understanding the implications of aggregator configurations is critical for designing market mechanisms that improve both efficiency and equity.

\section{Conclusion and Future work}\label{sec:conclusion-future-work}
In this paper, we have extended the fair energy resource allocation problem to a multi-aggregator setting, where multiple aggregators interact within an electricity market. We prove that the strategic optimization faced by aggregators forms a quasiconcave game, and we demonstrate the existence of a Nash equilibrium and conjecture on the uniqueness of the converged Nash equilibrium. Our paper presents valuable insights into how aggregator market structures can promote market stability while realizing fairness and efficiency in resource allocation. Our theoretical framework, supported by simulations, illustrates how aggregators stabilize market dynamics, shape users' surplus/consumption distributions, and manage system fairness-efficiency trade-offs, particularly for small-scale users.

The simulations further validated the theoretical results, showing how different market configurations and fairness objectives influence small users' surplus and consumption distribution. The introduction of large users was shown to significantly impact the surplus of small users, which is mitigated when aggregators are introduced. Aggregators operating under proportional fairness or social welfare objectives were seen to improve the outcomes for smaller users, balancing fairness and efficiency. Furthermore, the simulations demonstrated how the number of aggregators and their fairness considerations shape the equilibrium behavior of the market.






\bibliographystyle{IEEEtran} 
\bibliography{Reference}

\begin{thebibliography}{10}
\providecommand{\url}[1]{#1}
\csname url@samestyle\endcsname
\providecommand{\newblock}{\relax}
\providecommand{\bibinfo}[2]{#2}
\providecommand{\BIBentrySTDinterwordspacing}{\spaceskip=0pt\relax}
\providecommand{\BIBentryALTinterwordstretchfactor}{4}
\providecommand{\BIBentryALTinterwordspacing}{\spaceskip=\fontdimen2\font plus
\BIBentryALTinterwordstretchfactor\fontdimen3\font minus \fontdimen4\font\relax}
\providecommand{\BIBforeignlanguage}[2]{{%
\expandafter\ifx\csname l@#1\endcsname\relax
\typeout{** WARNING: IEEEtran.bst: No hyphenation pattern has been}%
\typeout{** loaded for the language `#1'. Using the pattern for}%
\typeout{** the default language instead.}%
\else
\language=\csname l@#1\endcsname
\fi
#2}}
\providecommand{\BIBdecl}{\relax}
\BIBdecl

\bibitem{FERC2022}
{FERC}, ``Participation of distributed energy resource aggregations in markets operated by regional transmission organizations and independent system operators,'' https://www.govinfo.gov/content/pkg/FR-2021-03-30/pdf/2021-06089.pdf, 20201.

\bibitem{burger2017review}
\BIBentryALTinterwordspacing
S.~Burger, J.~P. Chaves-Ávila, C.~Batlle, and I.~J. Pérez-Arriaga, ``A review of the value of aggregators in electricity systems,'' \emph{Renewable and Sustainable Energy Reviews}, vol.~77, pp. 395--405, 2017. [Online]. Available: \url{https://www.sciencedirect.com/science/article/pii/S1364032117305191}
\BIBentrySTDinterwordspacing

\bibitem{chen2023competitive}
\BIBentryALTinterwordspacing
C.~Chen, A.~S. Alahmed, T.~D. Mount, and L.~Tong, ``Competitive der aggregation for participation in wholesale markets,'' in \emph{Hawaii International Conference on System Sciences}, 2023. [Online]. Available: \url{https://scholarspace.manoa.hawaii.edu/server/api/core/bitstreams/fb77470a-ef2a-43a3-9ffc-61906dd13000/content}
\BIBentrySTDinterwordspacing

\bibitem{moret2019energy}
F.~Moret and P.~Pinson, ``Energy collectives: A community and fairness based approach to future electricity markets,'' \emph{IEEE Transactions on Power Systems}, vol.~34, no.~5, pp. 3994--4004, 2019.

\bibitem{zhang2015competition}
B.~Zhang, R.~Johari, and R.~Rajagopal, ``Competition and coalition formation of renewable power producers,'' \emph{IEEE Transactions on Power Systems}, vol.~30, no.~3, pp. 1624--1632, 2015.

\bibitem{li2024balancing}
J.~Li, M.~Motoki, and B.~Zhang, ``Balancing fairness and efficiency in energy resource allocations,'' in \emph{IEEE Conference on Decision and Control}, 2024.

\bibitem{yang2021optimal}
Y.~Yang, G.~Hu, and C.~J. Spanos, ``Optimal sharing and fair cost allocation of community energy storage,'' \emph{IEEE Transactions on Smart Grid}, vol.~12, no.~5, pp. 4185--4194, 2021.

\bibitem{fornier2024fairness}
Z.~Fornier, V.~Lecl{\`e}re, and P.~Pinson, ``Fairness by design in shared-energy allocation problems,'' \emph{arXiv preprint arXiv:2402.00471}, 2024.

\bibitem{kirschen2004fundamental}
D.~Kirschen and G.~Strbac, \emph{\BIBforeignlanguage{English}{Fundamentals of Power System Economics}}.\hskip 1em plus 0.5em minus 0.4em\relax United Kingdom: John Wiley \& Sons Ltd, 2004.

\bibitem{hobbs2001equilibrium}
B.~Hobbs, U.~Helman, and J.-S. Pang, ``Equilibrium market power modeling for large scale power systems,'' in \emph{2001 Power Engineering Society Summer Meeting. Conference Proceedings (Cat. No. 01CH37262)}, vol.~1.\hskip 1em plus 0.5em minus 0.4em\relax IEEE, 2001, pp. 558--563.

\bibitem{wei2014competitive}
E.~Wei, A.~Malekian, and A.~Ozdaglar, ``Competitive equilibrium in electricity markets with heterogeneous users and price fluctuation penalty,'' in \emph{53rd IEEE Conference on Decision and Control}, 2014, pp. 6452--6458.

\bibitem{rodriguez2021value}
R.~Rodr{\'\i}guez, M.~Negrete-Pincetic, N.~Figueroa, {\'A}.~Lorca, and D.~Olivares, ``The value of aggregators in local electricity markets: A game theory based comparative analysis,'' \emph{Sustainable Energy, Grids and Networks}, vol.~27, p. 100498, 2021.

\bibitem{bruninx2019interaction}
K.~Bruninx, H.~Pand{\v{z}}i{\'c}, H.~Le~Cadre, and E.~Delarue, ``On the interaction between aggregators, electricity markets and residential demand response providers,'' \emph{IEEE Transactions on Power Systems}, vol.~35, no.~2, pp. 840--853, 2019.

\bibitem{akorede2010ders}
\BIBentryALTinterwordspacing
M.~F. Akorede, H.~Hizam, and E.~Pouresmaeil, ``Distributed energy resources and benefits to the environment,'' \emph{Renewable and Sustainable Energy Reviews}, vol.~14, no.~2, pp. 724--734, 2010. [Online]. Available: \url{https://www.sciencedirect.com/science/article/pii/S1364032109002561}
\BIBentrySTDinterwordspacing

\bibitem{sarker2016optimal}
M.~R. Sarker, Y.~Dvorkin, and M.~A. Ortega-Vazquez, ``Optimal participation of an electric vehicle aggregator in day-ahead energy and reserve markets,'' \emph{IEEE Transactions on Power Systems}, vol.~31, no.~5, pp. 3506--3515, 2016.

\bibitem{contreras2017participation}
J.~E. Contreras-Ocana, M.~A. Ortega-Vazquez, and B.~Zhang, ``Participation of an energy storage aggregator in electricity markets,'' \emph{IEEE Transactions on Smart Grid}, vol.~10, no.~2, pp. 1171--1183, 2017.

\bibitem{xie2022information}
L.~Xie, T.~Huang, P.~Kumar, A.~A. Thatte, and S.~K. Mitter, ``On an information and control architecture for future electric energy systems,'' \emph{Proceedings of the IEEE}, vol. 110, no.~12, pp. 1940--1962, 2022.

\bibitem{mo2000fair}
J.~Mo and J.~Walrand, ``Fair end-to-end window-based congestion control,'' \emph{IEEE/ACM Transactions on networking}, vol.~8, no.~5, pp. 556--567, 2000.

\bibitem{low2002internet}
S.~H. Low, F.~Paganini, and J.~C. Doyle, ``Internet congestion control,'' \emph{IEEE control systems magazine}, vol.~22, no.~1, pp. 28--43, 2002.

\bibitem{pinto2011new}
T.~Pinto, H.~Morais, P.~Oliveira, Z.~Vale, I.~Pra{\c{c}}a, and C.~Ramos, ``A new approach for multi-agent coalition formation and management in the scope of electricity markets,'' \emph{Energy}, vol.~36, no.~8, pp. 5004--5015, 2011.

\bibitem{wolff2023dynamic}
T.~Wolff and A.~Nie{\ss}e, ``Dynamic overlapping coalition formation in electricity markets: An extended formal model,'' \emph{Energies}, vol.~16, no.~17, p. 6289, 2023.

\bibitem{lu2020fundamentals}
X.~Lu, K.~Li, H.~Xu, F.~Wang, Z.~Zhou, and Y.~Zhang, ``Fundamentals and business model for resource aggregator of demand response in electricity markets,'' \emph{Energy}, vol. 204, p. 117885, 2020.

\bibitem{chen2024wholesale}
C.~Chen, S.~Bose, T.~D. Mount, and L.~Tong, ``Wholesale market participation of deras: Dso-dera-iso coordination,'' \emph{IEEE Transactions on Power Systems}, 2024.

\bibitem{mathur2017optimal}
S.~P. Mathur, A.~Arya, and M.~Dubey, ``Optimal bidding strategy for price takers and customers in a competitive electricity market,'' \emph{Cogent Engineering}, vol.~4, no.~1, p. 1358545, 2017.

\bibitem{zhao2022strategic}
D.~Zhao, M.~Jafari, A.~Botterud, and A.~Sakti, ``Strategic energy storage investments: A case study of the caiso electricity market,'' \emph{Applied Energy}, vol. 325, p. 119909, 2022.

\bibitem{debreu1952social}
G.~Debreu, ``A social equilibrium existence theorem,'' \emph{Proceedings of the national academy of sciences}, vol.~38, no.~10, pp. 886--893, 1952.

\bibitem{osborne1994course}
M.~J. Osborne, \emph{A Course in Game Theory}.\hskip 1em plus 0.5em minus 0.4em\relax MIT Press, 1994.

\bibitem{berry2013economic}
R.~A. Berry, R.~Johari \emph{et~al.}, ``Economic modeling in networking: A primer,'' \emph{Foundations and Trends{\textregistered} in Networking}, vol.~6, no.~3, pp. 165--286, 2013.

\bibitem{cambini2008generalized}
A.~Cambini and L.~Martein, \emph{Generalized convexity and optimization: Theory and applications}.\hskip 1em plus 0.5em minus 0.4em\relax Springer Science \& Business Media, 2008, vol. 616.

\bibitem{li2017distributed}
P.~Li, H.~Wang, and B.~Zhang, ``A distributed online pricing strategy for demand response programs,'' \emph{IEEE Transactions on Smart Grid}, vol.~10, no.~1, pp. 350--360, 2017.

\bibitem{khezeli2017risk}
K.~Khezeli and E.~Bitar, ``Risk-sensitive learning and pricing for demand response,'' \emph{IEEE Transactions on Smart Grid}, vol.~9, no.~6, pp. 6000--6007, 2017.

\bibitem{patnam2021demand}
B.~S.~K. Patnam and N.~M. Pindoriya, ``Demand response in consumer-centric electricity market: Mathematical models and optimization problems,'' \emph{Electric Power Systems Research}, vol. 193, p. 106923, 2021.

\bibitem{agrawal2020disciplined}
A.~Agrawal and S.~Boyd, ``Disciplined quasiconvex programming,'' \emph{Optimization Letters}, vol.~14, no.~7, pp. 1643--1657, 2020.

\end{thebibliography}






\end{document}